\documentclass[prl,aps,twocolumn,amsfonts,showpacs,superscriptaddress,floatfix]{revtex4-1}

 \usepackage{epsfig} \usepackage{psfrag} \usepackage{amsmath}
 \usepackage{amssymb} \usepackage{color} \usepackage{graphicx}
 \usepackage{subfigure}

 \renewcommand{\vec}[1]{\mathbf{#1}}

\begin{document}

\title{Fermionic Superradiance in a Transversely Pumped Optical
  Cavity} \author{J. Keeling} \affiliation{SUPA,
School of Physics and Astronomy, University of St
  Andrews, St Andrews KY16 9SS, United Kingdom} \author{M. J. Bhaseen}
\affiliation{Department of Physics, King's College London, Strand, London WC2R 2LS, United Kingdom} \author{B. D. Simons}
\affiliation{University of Cambridge, Cavendish Laboratory, Cambridge
  CB3 0HE, United Kingdom}  \date{\today}

  \begin{abstract}
 Following the experimental realization of Dicke superradiance in Bose
 gases coupled to cavity light fields, we investigate the behavior of
 ultra cold fermions in a transversely pumped cavity. We focus on the
 equilibrium phase diagram of spinless fermions coupled to a single
 cavity mode and establish a zero temperature transition to a
 superradiant state.  In contrast to the bosonic case, Pauli blocking
 leads to lattice commensuration effects that influence
 self-organization in the cavity light field. This includes a sequence
 of discontinuous transitions with increasing atomic density and
 tricritical superradiance.  We discuss the implications for
 experiment.
 \end{abstract} 

 \date{\today}


 \maketitle

 {\em Introduction.}--- Experiments combining cold atomic gases with
 cavity quantum electrodynamics (QED) have led to pivotal developments
 in matter-light interaction. The use of Bose--Einstein condensates
 (BECs) allows precise control over the collective matter-light
 coupling, and permits access to the strong coupling regime
 \cite{Brennecke:Cavity,Colombe:Strong}.  This may be exploited for
 spectroscopy of many body systems
 \cite{Ritsch:Probing,Chen:Numstat,Brahms:MRI}, and to induce
 light-mediated interactions.

 Early theoretical work predicted that atoms in a cavity undergo a
 self-organization transition when pumped transversely
 \cite{Domokos:Collective,Nagy:SO,Nagy:Self}.  This was confirmed by
 experiments on thermal clouds and recently on a BEC
 \cite{Black:SO,Baumann:Dicke,Baumann:Dicke2}.  The latter also
 established equivalence to the superradiance transition in an
 effective Dicke model
 \cite{Dicke:Coherence,Hepp:Super,Wang:Dicke,Emary:Chaos,Dimer:Proposed,Nagy:Dicke}. These
 advances open the door to non-equilibrium and strongly correlated
 matter-light phenomena, including driven-dissipative phase
 transitions \cite{Brennecke:Real,Torre:Keldysh}, Mott insulator
 transitions in self-organized lattices
 \cite{Larson:MIRes,Larson:Stability} and cavity optomechanics
 \cite{Stamperkurn:Optoreview}.  They also provide a platform on which
 to explore frustrated spin models and glassy behavior in multimode
 cavities
 \cite{Gop:Emergent,Gopal:Atomlight,Gopal:Frus,Strack:Dicke,Muller:Charge}.
 For a review see \cite{Ritsch:RMP}.

The experimental realization of superradiance in BECs raises many
questions regarding the possible behavior of fermions in cavities.
Recent investigations have considered longitudinal pumping
\cite{Larson:Cold,Kanamoto:Opto}, cavity mediated pairing
\cite{Guo:QEDBCSBEC} and synthetic gauge fields \cite{Padhi:Cavity}.
In this manuscript we focus on the closest extension of recent
experiments \cite{Baumann:Dicke,Baumann:Dicke2} by coupling spinless
fermions to a single mode of a transversely pumped cavity.  At high
temperatures, where both fermions and bosons exhibit
Maxwell--Boltzmann statistics, it is evident that fermions will
exhibit a superradiance transition, as observed for thermal bosons
\cite{Black:SO}.  However, at low temperatures the Bose and Fermi
gases are expected to behave differently and numerous questions
arise. Does the self-organization transition survive for degenerate
fermions?  How does commensurability between the Fermi wavevector and
the self-consistent optical lattice affect self-organization?  Do new
phases exist, and what characterizes the transitions?

\begin{figure}[htpb]
 \includegraphics[clip=true,width=3.2in]{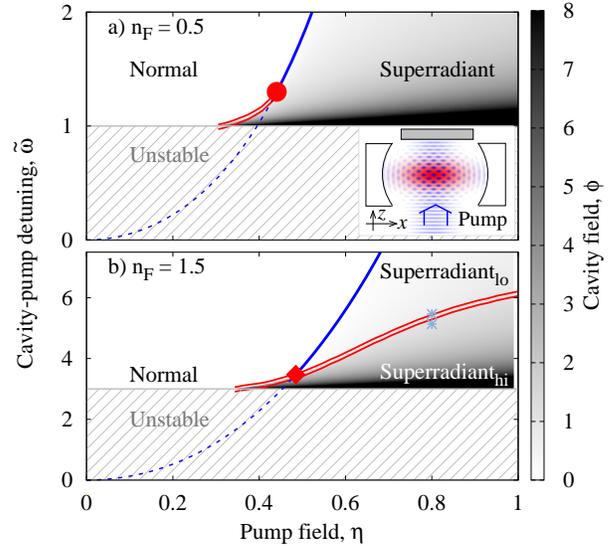}
 \caption{Equilibrium phase diagram of fermions in a transversely
   pumped cavity; see inset. As the pumping is increased there is a
   transition to a superradiant (SR) state, where the fermions
   spontaneously ``self-organize'' in the self-consistent light field.
   The panels correspond to partial filling of (a) the first and (b)
   second bands of the emergent lattice.  At large detuning the cavity
   field (grayscale) grows continuously above a critical pump field
   (solid blue line), whilst at smaller detuning the transition is
   discontinuous (double red line).  These first and second order
   boundaries join differently at different fillings; for $n_F
   \lesssim 1$ they meet at a tricritical point (circle) whilst at
   higher fillings there is a critical endpoint (diamond). The first
   order boundary in (b) corresponds to a liquid-gas type transition
   within the SR phase.  The unstable region $\tilde\omega<2 n_F$ will
   be stabilized by cavity loss; see text. The dashed line is the
   spinodal extension of the continuous line.  }
 \label{Fig:Eq}
 \end{figure}

 {\em The Model.}--- Inspired by
 Refs.~\cite{Black:SO,Baumann:Dicke,Baumann:Dicke2}, we consider
 spinless fermions coupled to a single mode of a cavity light field
 (forming a standing wave in the $x$-direction), and pumped by a
 transverse laser (in the $z$-direction); see inset of
 Fig.~\ref{Fig:Eq}.
We assume that the pump frequency $\omega_p$ is far detuned from the
atomic transition frequency $\omega_a$ so that absorption and
consequent heating via spontaneous emission can be neglected. After
eliminating the excited states of the atoms (see
e.g. \cite{Bhaseen:Noeqdicke}), one obtains an effective Hamiltonian
governing the interaction between the cavity light field and the
motional degrees of freedom:
 \begin{multline}
   \label{model}
   \hat{H} = \hbar \omega \hat\psi^\dagger \hat \psi
   + \sum_{\vec{k}} \left[
     \frac{\hbar^2 k^2}{2m} 
     \hat c^\dagger_{\vec{k}} \hat c^{}_{\vec{k}} 
     -
     \frac{g^2}{4 \Delta_a} \hat \psi^\dagger \hat \psi
     \sum_{s}
     \hat c^\dagger_{\vec{k}} \hat c^{}_{\vec{k}+ 2 s \vec{q}_x}
   \right.
   \\\left.
     -
     \frac{g \Omega}{4\Delta_a} (\hat \psi + \hat \psi^\dagger)\!
     \sum_{s,s^\prime}\!
     \hat c^\dagger_{\vec{k}} \hat c^{}_{\vec{k}+ s \vec{q}_z  + s^\prime \vec{q}_x} 
     \!\!\!-\!
     \frac{\Omega^2}{4\Delta_a} \!\sum_s\!
     \hat c^\dagger_{\vec{k}} \hat c^{}_{\vec{k}+ 2 s \vec{q}_z}
   \right]
 \end{multline}
 where $\Delta_a\equiv \omega_a-\omega_p$ and $s,s^\prime\in \pm 1$
 denotes the forward and backward components of the standing
 waves. Here, $\hat{c}_\vec{k}$ is an annihilation operator for a
 spinless fermion with mass $m$ and wavevector $\vec{k}$, and
 $\hat{\psi}$ is a bosonic annihilation operator for cavity
 photons. Eq.~(\ref{model}) is written in the rotating frame of the
 pump so $\omega\equiv\omega_c-\omega_p$ is the cavity-pump
 detuning. To ensure efficient scattering between the pump and the
 cavity, the cavity frequency $\omega_c$ is assumed to be close to
 $\omega_p$.  The cavity and pump light fields thus have approximately
 equal wavelengths with $|\vec{q}_x| = |{\vec q}_z|=q$. Scattering
 between the pump and the cavity involves the cavity-atom coupling
 $g$, and the pump strength $\Omega$, and is described by the fourth
 term in Eq.~(\ref{model}).  The third (fifth) term corresponds to a
 second order process involving the absorption and emission of cavity
 (pump) photons.  For bosons, there are limits where one may truncate
 the number of ${\bf k}$-states; when truncated to $\vec{k}=(0,0)$ and
 $\vec{k}\in (\pm q, \pm q)$, the Hamiltonian maps on to an effective
 Dicke model describing two-level systems coupled to light
 \cite{Dimer:Proposed,Baumann:Dicke,Nagy:Dicke,Bhaseen:Noeqdicke}.
 For fermions, Pauli blocking generally precludes truncation and so we
 must consider the higher ${\bf k}$-states.

In a ``self-organized'' state, the cavity light field develops an
expectation $\langle \hat{\psi} \rangle \neq 0$, and the superposition
of the pump and cavity fields forms a 2D lattice with reciprocal
lattice vector $(q,q)$.  We introduce dimensionless units by measuring
atomic energies in units of the recoil energy $E_R \equiv\hbar^2 q^2 /
2m$. Wavevectors (and lengths) are measured in units of the magnitude
of the reciprocal lattice vector $\sqrt{2} q$ so that the resulting
Brillouin Zone (BZ) has unit area.  We consider atoms confined to a 2D
layer in the $x,z$ plane and thus define the filling fraction $n_F
\equiv N/N_l$ as the number of atoms per lattice site; $N_l =
2q^2\mathcal{A}/(2\pi)^2$ for a real space area $\mathcal{A}$.  From
Eq.~(\ref{model}) it is natural to introduce dimensionless pump and
cavity fields via $\eta^2 = \Omega^2/4\Delta_a E_R$ and $\phi^2 = g^2
\langle \hat \psi \rangle^2 /4 \Delta_a E_R$.

 {\em Equilibrium Phase Diagram.}--- We begin by determining the
 equilibrium phase diagram for the Hamiltonian (\ref{model}). Although
 this neglects cavity losses, key features will survive in this limit
 \cite{Baumann:Dicke,Baumann:Dicke2}.  We treat the cavity mode in
 mean-field theory, which is exact in the thermodynamic limit $N,
 \mathcal{A} \to \infty$ \footnote{The exactness of mean field theory
   in the $N, \mathcal{A} \to \infty$ limit may be seen using a
   path-integral representation of $f$.  After eliminating atomic
   fields, fluctuation corrections to the mean-field saddle point are
   suppressed by $N/\mathcal{A}^2$, and so vanish in the thermodynamic
   limit, as also occurs in the bosonic version of this
   model~\cite{Piazza2013}.}.  Considering the dimensionless free
 energy density $f=F/(E_R N_l)$, we find:
 \begin{equation}
   f
   =
   \tilde{\omega} |\phi|^2 
   - {\tilde\beta}^{-1}   \int_{BZ} \!\!\!\! d^2k \sum_{n}
   \ln \left[ 1 + e^{-\tilde\beta(\epsilon_{\vec{k},n} - \mu)} \right]+\mu n_F,
 \label{free}
 \end{equation}
where $\tilde{\omega} \equiv \omega(4\Delta_a / g^2 N_l)$ is a
dimensionless cavity-pump detuning.  Here, $\epsilon_{\vec{k},n}$ is
the energy in the $n$th band, found by diagonalizing the atomic part
of Eq.~(\ref{model}).  Both $\epsilon_{\vec{k},n}$ and $\mu$ are in
units of $E_R$, and $\tilde\beta\equiv E_R/k_BT$.  To ensure that
$\mu$ is unambiguously defined, even for filled bands, we work at a
low non-zero temperature, $k_BT = 0.05 E_R$, throughout.  Minimization
of $f$ at fixed $n_F$ yields Fig.~\ref{Fig:Eq}.

 Figure~\ref{Fig:Eq} shows two fillings, characteristic of a partially
 filled first band (panel a, $n_F=0.5$) and a partially filled second
 band (panel b, $n_F=1.5$).  In both cases, two phases exist.  At low
 $\eta$ there is a normal state with $\phi=0$.  At large $\eta$,
 $\phi\neq 0$ and this state is labelled ``superradiant'' (SR) by
 analogy with the Dicke model terminology
 \cite{Hepp:Super,Wang:Dicke,Emary:Chaos,Dimer:Proposed,Nagy:Dicke}. It
 is also ``self-organized'' as the fermions are arranged in a
 self-consistent optical lattice.

 {\em Landau Theory.}--- As found in the bosonic case
 \cite{Baumann:Dicke,Baumann:Dicke2}, the normal-SR transition is
 second order at high $\eta$ and $\tilde\omega$. However, for the
 partially filled first band, on decreasing $\eta$, a tricritical
 point occurs beyond which the transition becomes first order.  This
 can be understood via a Landau expansion, $f = f_0 +
 a(\tilde{\omega},\eta,n_F) \phi^2 + b(\eta,n_F) \phi^4 + c(\eta,n_F)
 \phi^6$, where pump-cavity phase-locking ensures $\phi\in {\mathbb
   R}$.  Taking $c>0$ for stability, three types of behavior occur
 depending on the sign of $b$ \cite{Chaikin:Book}.  For $b>0$ a
 continuous transition occurs at $a=0$, whilst a first order
 transition occurs at $a=b^2/4c$ when $b<0$.  These transitions meet
 at a tricritical point at $a=b=0$.  In the vicinity of the
 tricritical point the critical exponent describing the onset of the
 cavity field $\phi$ changes from $\beta=1/2$ to $\beta=1/4$.  At low
 $\eta$, $b(\eta,n_F)<0$ and so the boundary becomes first order in
 Fig.~\ref{Fig:Eq} (a).

 The physical origin of this discontinuous transition is reminiscent
 of the Larkin--Pikin mechanism \cite{Larkin:First}, where coupling to
 an additional degree of freedom drives a transition first
 order. Here, the order parameter $\phi$ couples to density waves of
 the atomic system.  The linear coupling to the
 $\cos(x/\sqrt{2})\cos(z/\sqrt{2})$ density wave described by the
 fourth term in Eq. (\ref{model}) yields a continuous transition.
 However, the quadratic coupling to the $\cos(\sqrt{2}x)$ density wave
 in the third term in Eq.~(\ref{model}) drives $b<0$ at small $\eta$,
 as follows from second order perturbation theory.  Note that this
 mechanism is distinct from the Brazovskii scenario for driving the
 self-organization transition first order in multimode cavities
 \cite{Gop:Emergent,Gopal:Atomlight}.

 The behavior described so far persists while only the first band is
 filled.  Richer behavior occurs when higher bands start to be filled;
 see Fig.~\ref{Fig:Eq} (b). Here, there is no tricritical point and
 the second order line terminates at a critical endpoint
 \cite{Chaikin:Book}. The SR phase is now divided by a first order
 transition, separating high and low $\phi$ regions.  This is
 analogous to a liquid-gas transition, and the two phases are
 connected by a trajectory at lower $n_F$.  Within Landau theory this
 corresponds to terms beyond $\phi^4$ being negative.  We will return
 to the physical origin of this transition later in the manuscript.

 {\em Unstable Regions.}--- As indicated in Fig.~\ref{Fig:Eq}, when
 $\tilde{\omega} < 2 n_F$, $f$ is unbounded from below.  This reflects
 the form of $f$ at large $\phi$, when the atoms are trapped in deep
 minima of the cavity optical lattice. Here, the leading contribution
 to the atomic energy is $\epsilon_{\vec{k}, n} \sim - 2 \phi^2$ and
 so $f \sim (\tilde{\omega} - 2 n_F) \phi^2$.  The unstable region
 exists even at low densities, where Pauli blocking can be ignored,
 and so it is also relevant for bosons.  Such a situation has been
 discussed for bosons in Ref.~\cite{Liu:Light}.  However, as we argued
 in Ref.~\cite{Bhaseen:Noeqdicke}, this instability will be replaced
 by dynamical attractors in the presence of cavity losses.

 {\em Continuous Phase Boundaries.}--- Where the boundaries are
 continuous, the $\eta, n_F$ dependence of the critical detuning
 $\tilde{\omega}$ can be obtained from the vanishing quadratic Landau
 coefficient.  This has contributions from the first and third terms
 in Eq.~(\ref{model}).  Using second order perturbation theory,
 $a(\tilde{\omega}, \eta, n_F) = \tilde{\omega} + \chi(\eta, n_F)$,
 where
 \begin{equation}
   \label{eq:1}
   \chi(\eta,n_F) = 
   2 \eta^2\!
   \int_{BZ} \!\!\!\!\!d^2k
   \sum_{ij} 
   n_F(\epsilon_{\vec{k},i})
   \frac{|\langle u_{\vec{k},i} | \hat{M} | u_{\vec{k},j} \rangle|^2}
   {\epsilon_{\vec{k},i} - \epsilon_{\vec{k},j}},
 \end{equation}
 is a dimensionless atomic susceptibility. Here,
 $\epsilon_{\vec{k},i}$ and $|u_{\vec{k},i} \rangle$ are the atomic
 energies and eigenstates evaluated in the absence of the cavity
 field. The pump-cavity scattering represented by the third term in
 Eq.~(\ref{model}) corresponds to $\hat{M} = 4 \cos(x/\sqrt{2})
 \cos(z/\sqrt{2})$ in the position basis.  At $\phi=0$ the
 wavefunctions $\langle x,z|u_{\vec{k},i} \rangle$ factorize into
 plane-waves in the $x$ (cavity) direction and Mathieu functions
 \cite{Whittaker:Modern} in the $z$ (pump) direction due to the pump
 lattice. The phase boundary occurs at $\tilde{\omega} = - \chi$; for
 parameters where the boundary turns first order, this becomes the
 spinodal line \cite{Chaikin:Book} as shown by the dashed lines in
 Fig.~\ref{Fig:Eq}.

 In the limits of low and high pump field, analytical results for the
 boundaries may be obtained. For $\eta\rightarrow 0$ the Mathieu
 functions are plane waves and one finds $\chi=- 4\pi \eta^2
 (1-\Theta(\gamma)\sqrt{|\gamma|})$ where $\Theta(\gamma)$ is a
 Heaviside step function and $\gamma=1-4 n_F/\pi$.  The sharp
 threshold reflects Pauli blocking: at high densities the
 susceptibility saturates due to states deep within the Fermi surface
 (FS) not contributing.  The threshold occurs when $n_F=\pi
 k_F^2=\pi/4$ or $2k_F=1$. i.e. when the FS diameter matches the
 scattering wavevector represented by the third term in
 Eq.~(\ref{model}). Note however that at low $\eta$ the phase boundary
 is actually first order, so $\tilde{\omega} =- \chi$ is a spinodal
 line.  The $\eta^2$ dependence is evident from the dashed line in
 Fig.~\ref{Fig:Eq}. In the low density and low pump field limit this
 corresponds to $\tilde{\omega} = 8 \eta^2 n_F$, which is where the
 Dicke approximation would erroneously predict a transition.  As
 $\eta\rightarrow\infty$, the Mathieu functions are very localized so
 the dispersion in $k_z$ is flat, while remaining quadratic in $k_x$.
 The bands induced by the pump are well separated and so only the
 lowest band need be considered, yielding $\chi = 16 \eta^2 \ln \left|
 (1-n_F)/(1+n_F) \right|$.  The divergence at $n_F=1$ is a consequence
 of FS nesting. For a flat $z$ dispersion the FS is delimited by
 $|k_z|<1/\sqrt{2}, |k_x|< n_F/2\sqrt{2}$, and the cavity-pump
 scattering induces atomic scattering, $k_x \to k_x \pm 1/\sqrt{2}$ at
 a nesting wavevector for $n_F=1$.  At finite $\eta$ the logarithmic
 singularity is softened by imperfect nesting but a peak remains
 unless $\eta\ll 1$; we will return to this below.

 \begin{figure}[htpb]
 \includegraphics[clip=true,width=3.2in]{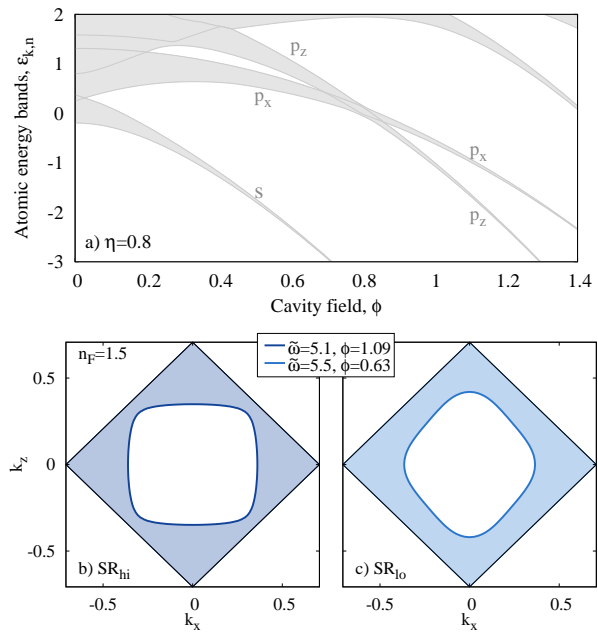}
\caption{(a) Atomic bands versus $\phi$ for $\eta=0.8$. When filling
  the 2nd or higher bands, $f(\phi)$ is non-monotonic, thus allowing
  first order liquid-gas type transitions in the SR phase. (b) and (c)
  illustrate a distortion of the FS on crossing the liquid-gas
  boundary; the values of $\tilde{\omega}$ are indicated by crosses in
  Fig.~\ref{Fig:Eq} (b). The FS delimits a partially filled second
  band and is shown in a reduced zone scheme. Occupied states are
  shaded.}
\label{Fig:Pom}
\end{figure}

{\em Liquid--Gas Transition.}--- As noted earlier, for partial filling
of the second (or higher) bands, a liquid--gas type transition
exists. The origin of this transition is revealed by the dispersion of
the self-consistent bands with cavity field $\phi$, as shown in
Fig.~\ref{Fig:Pom} (a).  There is a clear crossing of the $p_x$- and
$p_z$-like bands when $\phi\simeq \eta$; the bands are labelled by the
symmetry of the Wannier orbitals at large $\phi$.  This crossing leads
to a kink in the atomic free energy at $\phi\simeq\eta$, and a maximum
in $f(\phi)$ separating two minima.  This yields a discontinuous jump
from a high field state (${\rm SR}_{\rm hi}$) with $\phi\gtrsim \eta$
to a low-field state (${\rm SR}_{\rm lo}$) with $\phi\lesssim \eta$,
accompanied by a distortion of the FS; see Figs.~\ref{Fig:Pom} (b) and
(c). More generally, the dispersion of the higher bands is
non-monotonic in $\phi$ and further band crossings can occur.  Filling
such bands leads to a non-monotonic $f(\phi)$, and hence additional
liquid--gas type transitions.

\begin{figure}[htpb]
\includegraphics[clip=true,width=3.2in]{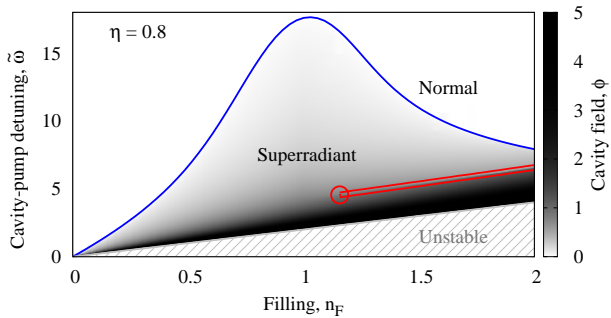}
\caption{Equilibrium phase diagram as a function of $n_F$ at
  $\eta=0.8$. The lines and the grayscale are as in
  Fig.~\ref{Fig:Eq}. A critical point (empty circle) terminates the
  liquid-gas like transition at $n_F\gtrsim 1$.  The second order
  normal-SR transition shows a peak near $n_F\simeq 1$ reflecting
  nesting; see text.}
\label{Fig:Trans}
\end{figure}

{\em Evolution with Filling.}--- Further insight into the liquid-gas
transition is obtained from the phase diagram as a function of $n_F$
at fixed $\eta$; see Fig.~\ref{Fig:Trans}. There is a critical point
within the SR phase for $n_F\gtrsim1$, beyond which a liquid-gas
boundary extends.  This corresponds to the point at which a sufficient
fraction of the 2nd band is filled in order to introduce the required
local minima in $f(\phi)$.  Also visible in Fig.~\ref{Fig:Trans} is a
peak in the second order boundary $\tilde{\omega}=-\chi$ near $n_F=1$.
As $\eta$ increases, this evolves into the logarithmic singularity of
$\chi$ at $n_F=1$, reflecting the FS nesting discussed above.

{\em Non-Equilibrium, Linear Stability.}--- Thus far we have provided
a detailed analysis of the equilibrium properties of
Eq.~(\ref{model}).  However, as discussed in
Refs.~\cite{Keeling:Collective,Bhaseen:Noeqdicke}, a finite cavity
decay rate $\kappa$ can significantly change the phase diagram.  In a
driven-open system, the phase diagram is found by determining the
stable attractors of the dynamics, not by minimizing the free
energy. Nonetheless, extrema of $f$ correspond to stationary points of
the dynamics in the limit $\kappa\rightarrow 0$.  The open system thus
inherits key features from its equilibrium counterpart
\cite{Baumann:Dicke,Baumann:Dicke2}. We next discuss which features
are robust to non-zero $\kappa$, and which aspects require further
investigation.

A feature that will survive when $\kappa\neq 0$ is the boundary at
which the normal state becomes unstable. This can be calculated by
finding when the normal mode frequencies have negative imaginary parts
\cite{strogatz94}. Following the approach used for the bosonic system
\cite{Dimer:Proposed,Bhaseen:Noeqdicke} the boundary of stability is
given by $(\tilde \kappa^2 + \tilde{\omega}^2)/\tilde{\omega} = -
\chi$, where $\chi$ is the atomic susceptibility given by
Eq.~(\ref{eq:1}) and $\tilde\kappa\equiv 4\kappa\Delta_a/g^2N_l$. In
the limit $\kappa\rightarrow 0$ one recovers the equilibrium result
$\tilde\omega=-\chi$ discussed above. This provides a direct link
between the equilibrium and non-equilibrium phase boundaries, as found
for the open Dicke model
\cite{Dimer:Proposed,Baumann:Dicke,Nagy:Dicke}.  Unfortunately, a
quantitative discussion of the fate of the first order boundaries is
more challenging in the open fermionic system. Nonetheless, the
existence of competing local minima in the equilibrium phase diagram
suggests that multiple dynamical attractors may survive in the open
limit. Likewise, determining the fate of the unstable region when
$\kappa\neq 0$ is difficult because it hinges on the long time
dynamics of the fermionic system. This could potentially involve limit
cycles, fixed points and chaotic attractors \cite{strogatz94}. Indeed,
the answer to the analogous question in the Dicke model
\cite{Bhaseen:Noeqdicke} demonstrates both superradiant phases and
limit cycles.  It would be interesting to explore this in more detail,
both theoretically and experimentally.

{\em Low Density Limit.}--- As noted above, at low pump field, the
normal-SR boundary always becomes first order, even as $n_F \to 0$.
At low densities, the integral in Eq.~(\ref{free}) simplifies, and one
finds that $b(\eta,n_F)/n_F \to - 1/2 + \mathcal{O}(\eta^2)$.  In this
low density limit, $k_F\ll q$, this fermionic result also applies to
bosons. Notably, this first order transition is absent from previous
studies of bosons in single mode cavities. In particular, this
discontinuous transition is lost when momentum states are truncated to
yield the effective Dicke model
\cite{Dimer:Proposed,Baumann:Dicke,Nagy:Dicke,Bhaseen:Noeqdicke}; the
transition involves hybridization with states $(\pm 2q,0)$ outside
this basis. In the cases where higher momentum states have been
considered \cite{Bhaseen:Noeqdicke,Piazza2013}, this first order
behavior was not captured due to focusing on the susceptibility and
second order boundaries.

{\em Conclusions.}--- We have explored the phase diagram of ultra cold
fermions in a transversely pumped cavity. In contrast to a BEC, the
interplay of the Fermi wavevector with the wavelength of the cavity
field leads to a rich dependence on the filling fraction.  We have
established distinct superradiance transitions whose character
reflects the impact of Pauli blocking and lattice
commensuration. Unlike the Dicke model, the phase boundary turns first
order at low pump field.  In addition to signatures in the cavity
light field, the measurable consequences include FS distortions and an
enhanced susceptibility near unit filling.  This study provides a
basis for future experimental and theoretical research including the
nature of dynamical attractors in the driven-dissipative system, the
impact of fermion interactions, and the behavior in multimode
geometries.

{\em Acknowledgments.}--- We thank T. Esslinger for discussions and
B. L. Lev for helpful comments on the manuscript. MJB and BDS
acknowledge EPSRC EP/J017639/1 and MJB thanks the Thomas Young Centre.
JK acknowledges support from EPSRC programme ``TOPNES''
(EP/I031014/1), EPSRC (EP/G004714/2), and hospitality from KITP Santa
Barbara.  This research was supported in part by the National Science
Foundation under Grant No. NSF PHY11-25915.

\end{document}